\documentclass[twocolumn,aps,showpacs,prb,tightenlines,amsmath,amssymb]{revtex4}
\usepackage{graphicx}% Include figure files
\usepackage{bm}% bold math
\usepackage{amssymb}
\usepackage{amsmath}
\usepackage{colordvi} 
\usepackage{color}

\begin{document}

\title{Unique Electron Spin Relaxation Induced by Confined Phonons in
  Nanowire-Based Quantum Dots}

\author{Y.\ Yin}
\author{M.\ W.\ Wu}
\thanks{Author to whom correspondence should be addressed}
\email{mwwu@ustc.edu.cn.}
\affiliation{Hefei National Laboratory for Physical Sciences at
  Microscale and Department of Physics, 
University of Science and Technology of China, Hefei,
  Anhui, 230026, China}

\date{\today}

\begin{abstract}
  Electron spin relaxation in nanowire-based quantum dots induced by confined
  phonons is investigated theoretically.  Due to the one-dimensional nature of the
confined phonons, the van Hove singularities of the confined
 phonons and the zero of the form factor of the electron-phonon 
coupling can lead to unique features of the spin relaxation rate.
Extremely strong spin relaxation can be obtained at the van Hove
  singularity. Meanwhile the spin relaxation rate can also
 be greatly suppressed at the
zero of the form factor. This unique feature indicates the flexibility
of nanowire-based quantum dots in the manipulation of spin states. It also
 offers a way to probe the property of the confined phonons.
\end{abstract}


\maketitle

Electron spin relaxation in semiconductor quantum dots (QDs) has been an
important problem due to the proposed application of spin states in QDs as
qubits for quantum computation.\cite{PhysRevA.57.120, RevModPhys.79.1217} QDs
with the spin-orbit coupling are of particular interest since they enable
efficient manipulations of spin with electric field.\cite{prl.91.126405,
  nature.427.50, prb.74.165319}  Typical III-V semiconductor QDs are either
self-assembled ones or fabricated by confining electrons in quantum wells by
electrodes, where bulk phonons in substrates play an important
role.\cite{prl.93.016601, prb.77.035323, PhysRevB.61.12639,
  PhysRevB.66.161318,jlcheng} For these QDs, the acoustic bulk phonons in
conjunction with the spin-orbit coupling serve as the main source for spin
relaxation in low-temperature regime.\cite{prb.77.035323} Recently, QDs based
on III-V compound semiconductor nanowires were fabricated.\cite{bjork:1058,
  bjork:1621, nilsson:163101} The nanowires are perpendicular to the substrate,
making the bulk phonons in substrate less important than the confined phonons in
nanowires.\cite{ohlsson:3335, shtrikman:2009}  The regular structure of the
nanowires results in the quasi-one-dimensional confined phonons, which lead to
novel properties in optical absorption and transport for the nanowire-based
QDs.\cite{PhysRevLett.99.087401, prl.104.036801} In this Report, we show that
the confined phonons lead to unique behaviors of the spin relaxation in elongate
QDs embedded in nanowires.

We focus here on a single-electron elongate QD embedded in InAs [001]
cylindrical nanowire with radius $R$ in the presence of an external magnetic
field $B$ along the wire. We model the QD by an anisotropic harmonic potential
$V_c(r, z) = \frac{1}{2} m^{\ast}\omega^2_0 r^2 + \frac{1}{2} m^{\ast}
\omega^2_z z^2$ with $z$-axis along the wire and $\omega_0 \gg \omega_z$ so that
only the lowest electron subband in the radial direction is needed.\cite{comment_rashba} $m^{\ast}$
is the effective mass of the electron. Thus the size of the QD is decided by the
diameter $d_0=\sqrt{\hbar \pi / m^{\ast} \omega_0}$ in radial direction and the
length $d_z = \sqrt{\hbar \pi / m^{\ast} \omega_z}$ in axial direction. The
spin-orbit coupling in the QD is described by the Rashba term $H_{\text{SO}} =
\frac{\gamma}{\hbar} \sigma_y p_z$, where $\gamma = 3.0 \times
10^{-11}$~eV$\cdot$m is Rashba coupling constant\cite{grundler} and $\sigma_y$
is the Pauli matrix.

Due to the well separated energy levels of the QD, the spin relaxation rate (SRR)
can be described by the scattering rate between the two lowest eigenstates $| i
\rangle$ and $| f \rangle$ with opposite spin orientations, which can be
calculated by the Fermi golden rule.\cite{PhysRevB.61.12639, PhysRevB.66.161318,
  jlcheng, prb.72.115326, prb.77.045328} Treating $| i \rangle$ and $| f
\rangle$ as the initial and final states, at zero temperature the SRR can be
written as
\begin{equation}\label{eq:srt}
  \Gamma_{fi} = \sum_{m \nu} \left. \frac{\left| M_{\nu}(\bm{q}_m) G_{fi}(\nu, \bm{q}_m) \right|^2}
    {\left|\partial_{\bm{q}_m} (\hbar \omega_{\nu}(\bm{q}_m)) \right|}\theta(\Delta \varepsilon_{fi})
  \right|_{\hbar \omega_{\nu}(\bm{q}_m) = \Delta \varepsilon_{fi}},
\end{equation}
where $\Delta \varepsilon_{fi} = \varepsilon_{i} - \varepsilon_{f}$ is the
energy splitting between the two states and $\theta(\varepsilon)$ is the
step function. $M_{\nu}(\bm{q}) G_{fi}(\nu, \bm{q})$ describes the matrix
element of the electron-phonon coupling 
with $G_{fi}(\nu, \bm{q})$ being the form factor,
which depends not only on the
electron wave function but also on the phonon eigenmode for confined
phonons. $\hbar
\omega_{\nu}(\bm{q})$ is the phonon spectrum. The summation $m$ 
is performed over the surface of the constant energy.

Poles in Eq.~(\ref{eq:srt}) correspond to the van Hove singularities. For 
bulk phonons, since the constant-energy surface is continuous, the SRR does not
diverge after the integration. However, for confined phonons, the
constant-energy surface reduces to discrete points, so the SRR can be
divergent if the energy splitting $\Delta \varepsilon$ matches the van Hove
singularities.  Similarly, the SRR can also be
suppressed by the zero of the form factor $G_{fi}(\nu, \bm{q})$. 
These features can be served as the fingerprints of the confined
phonons. We will show that these unique features of the SRR can be realized in typical
 nanowire-based QDs.\cite{bjork:1058, bjork:1621, nilsson:163101}
We mainly focus on the electron-phonon interaction due to the 
deformation potential coupling, since it is dominant 
for small semiconductor nanostructures.\cite{PhysRevLett.71.3577}

We calculate the confined phonons with isotropic elastic continuum model which
is widely used in the study of nanowires, carbon nanotubes and
nanoparticles.\cite{cleland_book,PhysRevLett.71.3577, suzuura_phonons_2002,%
  chassaing_raman_2009,yu_electronacoustic-phonon_1995,
  komirenko_renormalization_1998} The nanowire is modeled as an infinite
cylinder with the stress vanishing at the surface of the wire. Since we only
consider the lowest subband of the electron in the radial direction, due to the
conservation of the angular momentum, only phonon modes with zero angular
momentum can couple to electrons by the deformation-potential
coupling.\cite{PhysRevB.54.1494} These modes are usually referred as the dilatation
modes.\cite{cleland_book,yu_electronacoustic-phonon_1995,
  komirenko_renormalization_1998} 

\begin{figure}
  \includegraphics[width=7.5cm]{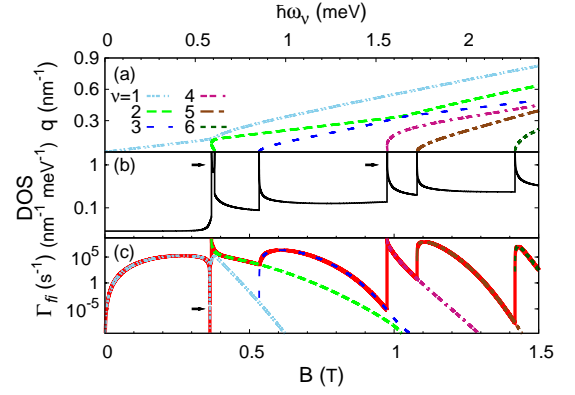}
  \caption{(Color online) SRR as a function of external magnetic field $B$ with
    QD diameter $d_0=12$~nm (shown in (c)). The corresponding phonon energy
    $\hbar \omega_{\nu}$ is given in the upper scale. (a) and (b) show the
    energy spectrum and DOS for dilatation phonon modes respectively. The red
    solid curve in (c) represents the total SRR. The contributions of each
    dilatation mode are plotted in (c) with different colors and line shapes.
    Curves with the same color and line shape in (a) and (c) correspond to the
    same phonon mode. The arrow in (c) indicates the dip induced
      by the zero of the form factor. The peaks in (b) indicated by the arrows
    correspond to the $q \ne 0$ van Hove singularities. The peak at $B=0.975$~T
    in fact contains two peaks too close to see due to the scale.}
  \label{fig:srt_b}
\end{figure}

We set the nanowire radius $R=15$~nm which is a typical radius for the
nanowires.\cite{shtrikman:2009} The other relevant parameters are the longitude
and transverse sound velocities which are chosen to be $v_L = 3830$~m/s and 
$v_T =2640$~m/s along the [001] direction.\cite{landolt_book}
The eigenmodes and spectrum
$\hbar \omega_{\nu}(q)$ of the confined phonons are calculated following
Refs.~\onlinecite{auld_book, stroscio:4670}. The calculated spectra of the first six
dilatation modes are plotted in Fig.~\ref{fig:srt_b}(a) and the corresponding
phonon density of the states (DOS) is shown in
 Fig.~\ref{fig:srt_b}(b). The peaks in the  DOS indicate
the van Hove singularities which correspond to the minima of each phonon
subband. It should be noted that there are two types of minima: minimum with
$q=0$ and $q\ne0$. Modes 3, 5 and 6 correspond to the 
first type. For modes 2 and 4, in addition to the first type, 
there are  minima with $q \ne 0$ 
($q = 0.063$~nm$^{-1}$ for mode 2 and $0.081$~nm$^{-1}$ for mode 4). 

The lowest two eigenstates $|i\rangle$ and $|f\rangle$ can be calculated by
treating the spin-orbit coupling as perturbation. Since only the lowest
electron subband is considered, the radial component of the electron wave
function is described by the cylindrical Gaussian function with width $d_0$. So the
unperturbed state $| n_z \sigma \rangle$ can be characterized by the quantum
number of the harmonic confinement along axial direction $n_z$ and electron spin
index $\sigma$. Up to the first-order perturbation, one has $| i \rangle = | 0
\uparrow \rangle + \mathcal{B}_{-} | 1 \downarrow \rangle$ and $| f \rangle = |0
\downarrow \rangle + \mathcal{B}_{+} | 1 \uparrow \rangle$, where
$\mathcal{B}_{\pm} = - \frac{\sqrt{2\pi}\gamma}{4 d_z}/(g \mu_B B \pm \hbar
\omega_z)$ with $g$ and $\mu_B$ being the $g$ factor of electron and Bohr
magneton, respectively.\cite{jlcheng}  Note that for the QD considered here, the energy
splitting induced by the spin-orbit coupling is much  smaller than the
Zeeman energy so it is adequate to have $\Delta \varepsilon = \left| g \mu_B B
\right|$.

Given the two eigenstates and the phonon eigenmodes, the matrix element of the
deformation-potential coupling $M_{\nu}(q) G_{fi}(\nu, q)$ can be
calculated\cite{PhysRevB.54.1494} with $q$ being the axial phonon wave
vector. The form factor is expressed as $G_{fi}(\nu, q) = I^d_z(q) I^d_{xy}(\nu,
q)$ with $I^d_z(q)$ and $I^d_{xy}(\nu, q)$ being the axial and radial
components. The axial component has the form $I^d_z(q)=d_z q e^{-(q d_z)^2/4
  \pi}$, which does not depend on the phonon mode index $\nu$. The radial
component can be written as
\begin{equation}
I^d_{xy}(\nu, q)= \chi^{(0)}_{\nu}(q) ( k_L / d_0 )^2 e^{A_L(\nu, q)},
\label{eq:form_def}
\end{equation}
with $k_{L,T} = {\hbar \omega_{\nu}(q)}/{(\hbar v_{L,T})}$ and $A_{L,T}(\nu, q)
= {(q^2-k^2_{L,T}) d_0^2}/{(4\pi)}$. $\chi^{(0)}_{\nu}(q)$ 
 is the coefficient in the
expression for the confined phonon eigenmode which are calculated numerically
following Refs.~\onlinecite{auld_book,stroscio:4670}. The corresponding
coefficient $M_{\nu}(q)$ can be expressed as $M_{\nu}(q) = |\mathcal{B}_{-} +
\mathcal{B}_{+}|^2 \mathcal{C}^d/\Delta \varepsilon$. $\mathcal{C}^d = \hbar^3
\Xi^2/( 16 D v_L )$ is a constant where the deformation-coupling strength
$\Xi=5.8$~eV and density $D=5900$~kg/m$^3$.\cite{landolt_book} Given the matrix
element $M_{\nu}(q) G_{fi}(\nu, q)$ and the phonon spectrum $\hbar
\omega_{\nu}(q)$, the SRR can be calculated by Eq.~(\ref{eq:srt}). In the
calculation, we set the size of the QD to be $d_z=50$~nm and
$d_0=12$~nm. $g=-14.7$.\cite{landolt_book}

The calculated SRR as a function of external magnetic field $B$ is plotted in
Fig.~\ref{fig:srt_b}(c). Since the SRR is the summation of the contributions
from each phonon mode, we plot them in the same figure with different
colors/line shapes for comparison. It can be seen that the SRR as a function of
$B$ can be separated into several regions. In each region, only one phonon mode
dominates. Thus the properties of individual phonon modes are crucial to the
SRR. This makes the SRR very sensitive to the magnetic field $B$. Sharp peaks
can be found in Fig.~\ref{fig:srt_b}(c) at $B=0.368$~T and $B=0.975$~T, 
whereas a dip exists at the position indicated by the arrow.

\begin{figure}
  \includegraphics[width=7.5cm]{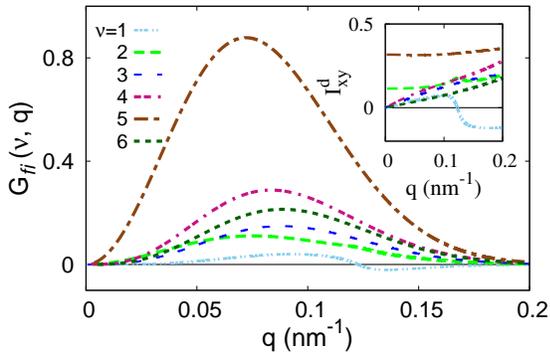}
  \caption{(Color online) Form factor $G_{fi}(\nu, q)$ for the first six
  dilatation modes versus wave vector $q$. 
Inset is for the radial component $I^d_{xy}(\nu,q)$. 
Curve with the same color and line shape as
that in Fig.~\ref{fig:srt_b} corresponds to the same
    dilatation mode.}
  \label{fig:form_factor}
\end{figure}

Let us first concentrate on the sharp peaks in the SRR. Comparing to the DOS in
Fig.~\ref{fig:srt_b}(b), one can see that these peaks correspond to the van Hove
singularities with $q \ne 0$. The SRR diverges at these singularities.  For the
van Hove singularities with $q = 0$, there is no divergence as the form factor
tends to 0 at $q=0$. Only broad peaks exist.  To show this, we plot the form
factor $G_{fi}(\nu, q)$ in Fig.~\ref{fig:form_factor}. Similar behavior also
exists in phonon satellites in the excitonic
absorption.\cite{PhysRevLett.99.087401} Note that this behavior is different
from the disk-shaped QDs embedded in the nanowires,\cite{PhysRevB.77.045434}
where the SRR diverges for singularities at $q=0$.  It is noted that the
divergence of the SRR at the van Hove singularities originates from the approach
of the Fermi golden rule Eq.~(\ref{eq:srt}), where the Markovian approximation
is implied. However, the divergence of the SRR means the memory effect should
not be neglected and one should apply the non-Markovian
approach\cite{PhysRevB.76.193312} to calculate the SRR around these
singularities. This will remove the divergence. Moreover, other mechanisms, such
as the phonon-phonon scattering, disorder, etc. can also remove the divergence
by broadening.  However, the peaks should still survive.

 Now we turn to the dip in the SRR indicated by the arrow in
  Fig.~\ref{fig:srt_b}(c). The dip results from the zero of the radial component
  $I^d_{xy}(\nu, q)$ of the form factor.\cite{comment_zero} From
 Fig.~\ref{fig:form_factor}, one can see that for phonon mode 1, $I^d_{xy}(\nu,
  q)$ has a zero point for $q \ne 0$. As a consequence, the deformation
  potential coupling of this mode vanishes at this point, making
 the SRR drop  to zero rapidly. It is further noted
 that according to Eq.~(\ref{eq:form_def}),
  the zero point of $I^d_{xy}(\nu, q)$ is just the zero of
 $\chi^{(0)}_{\nu}(q)$. So the dip in the SRR offers a way to probe the
confined phonons.  

In conclusion, we have calculated the SRR induced by confined acoustic phonons
in elongate QDs embedded in InAs [001] nanowires. The SRR is dominated by
individual confined phonon mode at different magnetic field region, which
results in a highly nonmonotonic magnetic field dependence.  Due to the
one-dimensional nature of the confined phonons, the SRR limited by the
spin-orbit coupling combined with the electron-phonon scattering can be
divergent at the van Hove singularities of the phonons, provided the form factor
does not go to zero at these singularities. Moreover, the 
zero of the form factor strongly suppresses the SRR, 
causing dips in the magnetic field dependence of the
SRR.  These features can be served as
the fingerprints of the confined phonons. It is also seen from our calculation
that the nanowire-based QDs can enable more flexible manipulations of spin
states.

This work was supported by the Natural Science Foundation of China under Grant
No.~10725417, the National Basic Research Program of China under Grant
No.~2006CB922005 and the Knowledge Innovation Project of Chinese Academy of
Sciences. One of the authors (MWW) would like to thank  Guido Burkard
for valuable discussions.


\begin{thebibliography}{33}
\bibitem{comment_rashba}  Strictly speaking, one has to use an axial
    asymmetric potential to describe the quantum dot in the nanowire, e.g., the
    center of the quantum dot is slightly displaced from the center of the
    nanowire, which induces the Rashba SOC.  But one can prove that, up to the
    first order, the contribution of the displacement to the spin relaxation
    vanishes. So it is adequate to use the axial symmetric model in the
    calculation.
\end{thebibliography}
\end{document}